\documentclass[prl,showpacs,noshowkeys,twocolumn]{revtex4}
\usepackage{graphicx}
\usepackage{bm}
\begin{document}
\title{Excluded volume causes integer and fractional plateaus\\ in colloidal ratchet currents}
\author{Pietro Tierno$^{1,2}$  \email{ptierno@ub.edu} and Thomas M. Fischer$^{3}$}
\affiliation{$^{1}$Estructura i Constituents de la Mat\'{e}ria, Universitat de Barcelona, Av. Diagonal 647, 08028 Barcelona, Spain\\
$^{2}$ Institut de Nanoci$\grave{e}$ncia i Nanotecnologia IN$^2$UB, Universitat de Barcelona, Barcelona, Spain\\
$^{3}$ Institute of Physics, Universit\"at Bayreuth, 95440 Bayreuth, Germany.}
\date{\today}
\begin{abstract}
We study the collective transport
of paramagnetic colloids driven above a magnetic bubble lattice
by an external rotating magnetic field.
We measure a direct ratchet current
which rises in integer and fractional steps with the field amplitude.
The stepwise increase is caused by excluded volume interactions between the particles,
which form composite clusters above the bubbles with mobile and immobile occupation sites.
Transient energy minima located at the interstitials between the bubbles
cause the colloids to hop from one composite cluster to the next with synchronous and period doubled modes
of transport.
The colloidal current may be polarized to make selective use of type up or type down interstitials.
\end{abstract}
\pacs{82.70.Dd, 05.60.Cd, 75.50.Gg}
\maketitle
The emergence of quantized steps in the current
of a driven system
is a fascinating phenomenon in condensed matter,
occurring in a broad range of systems such us in charge density waves~\cite{densitywave1,densitywave2},
in driven vortex lattices~\cite{vortex1,vortex2}, in sliding frictional surfaces~\cite{friction}
or in electronic tunneling~\cite{tunneling1,tunneling2}.
The impossibility of visualizing condensed matter quasi particles
has triggered the use
of alternative model systems
to unveil the basic mechanism leading to such transport
behaviour. Colloidal particles with accessible length-scale and dynamics,
represent a versatile model system with tuneable interactions~\cite{Model1,Model2,Model3}.
In particular, hard sphere interactions
creating excluded volume~\cite{excluded1,excluded2,excluded3} are relevant for the rheological properties of
colloidal dispersions~\cite{rheology1,rheology2} and they dominate the dynamics
near the colloidal glass transition~\cite{glass1,glass2}.
Here we show that the excluded
volume between paramagnetic colloids driven
above a magnetic bubble lattice causes
a series of discrete plateaus in the particle current, separated by steps where
some particles abruptly loose or gain mobility.
Integer plateaus result from particles moving
synchronously with the driving field,
while fractional plateaus arise from
nonlinear period doubling
with particles moving only every second cycle.
In contrast to many ratchet mechanisms~\cite{ratchet1,ratchet2,ratchet3} 
which operate under negligible particle-particle interactions, 
we introduce a colloidal ratchet where quantized transport phenomena arise 
due to excluded volume between the particles.\\
\begin{figure}[ht]
\begin{center}
\includegraphics[width=\columnwidth,keepaspectratio]{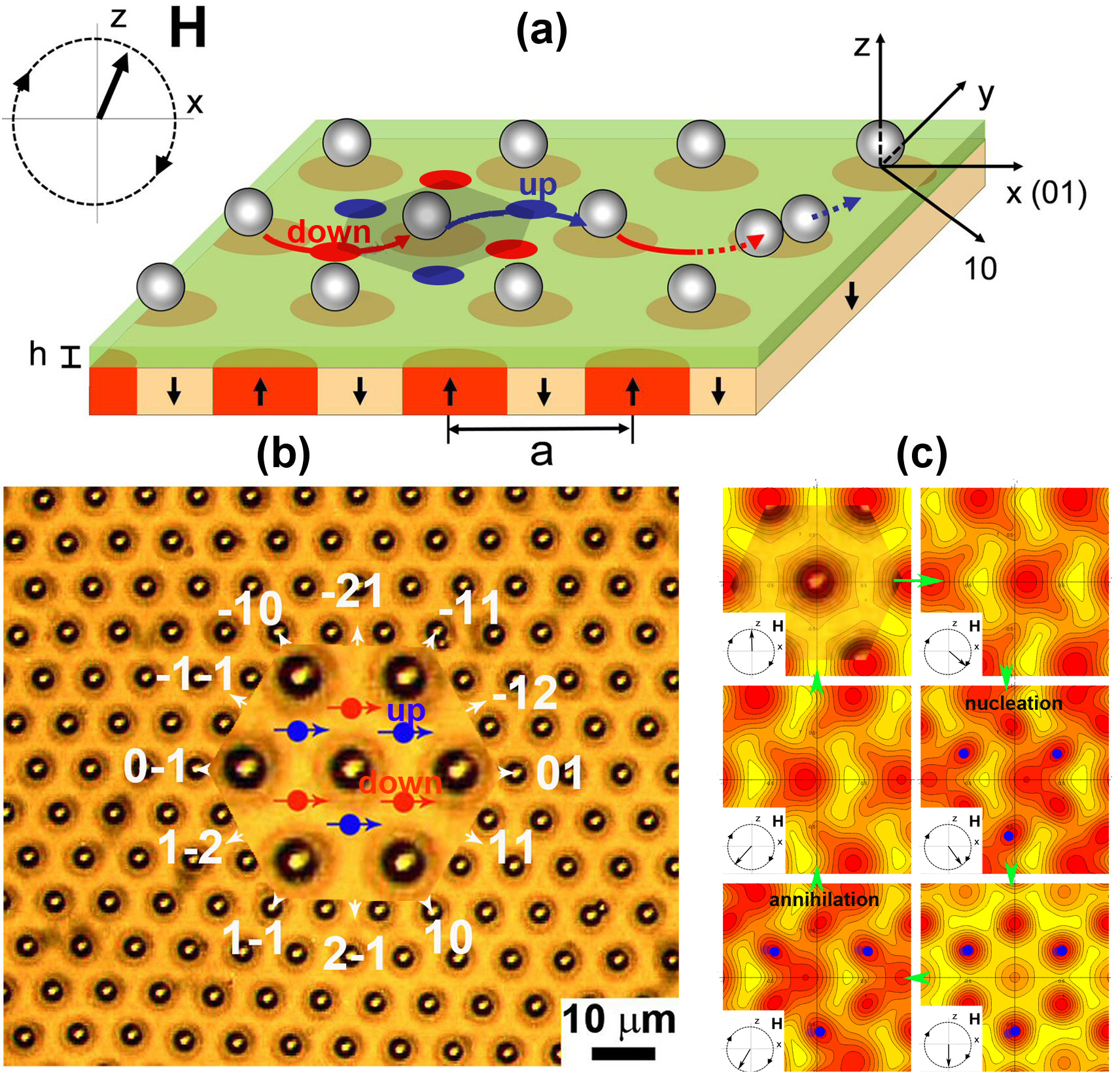}
\caption{(color online)(a) Schematic of the FGF film ($a=8.6 \, \mu m$)
covered with a polymer film ($h=1 \mu m$),
with one Wigner-Seitz unit cell shaded.
Potential particle paths between the bubbles
pass the type up (blue) and type down (red) interstitial regions.
(b) Polarization microscopy image of the FGF loaded with
one particle per unit cell. Crystal directions are indicated in white, type up
and type down interstitial wells are marked in blue
and in red, and nucleate at the beginning 
of the arrows, annihilate at the end. Central region has been magnified for clarity.
(c) Magnetic potential energy of a paramagnetic particle
during different phases of the applied field (inset).
Energy maxima are colored in yellow, minima in red.}
\label{figure1}
\end{center}
\end{figure}
We use monodisperse polystyrene paramagnetic
colloids dispersed
in deionized water and moving on top of
ferrite garnet film (FGF)
characterized by a triangular lattice of magnetic bubble domains~\cite{Bubble},
Fig.1(a). 
The FGF film exerts magnetic attraction 
to the paramagnetic colloids, and confines their motion in two dimensions.
A direct ratchet current
is obtained by modulating the heterogeneous stray field
of the FGF with an external field elliptically polarized in the $(x,z)$ plane,
$\bm{H}_{ext} \equiv (H_x\sin{(\omega t)},0,H_z\cos{(\omega t)})$,
more details are given in~\cite{EPAPS}.
Here $H_z$ indicates the component perpendicular to the FGF,
$H_x$ the parallel,
and $\omega$ the angular frequency.
Fig.1(c) illustrates how the potential energy landscape of a paramagnetic colloid is altered by the applied field
during one field cycle.
\begin{figure}[ht]
\begin{center}
\includegraphics[width=\columnwidth,keepaspectratio]{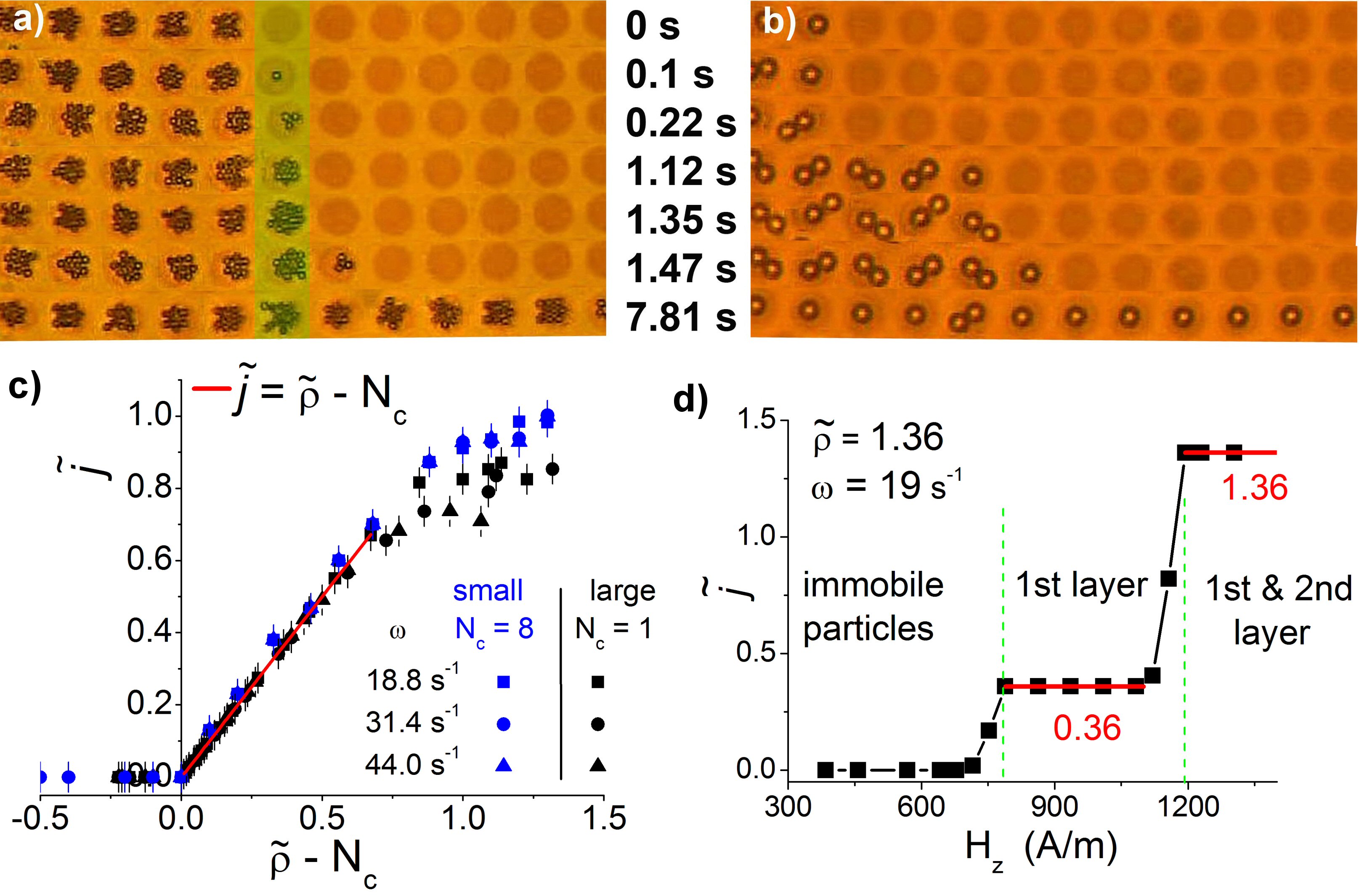}
\caption{(color online)(a) Time sequence of polarization microscopy images showing a row of bubbles subsequently filled
with paramagnetic particles of diameter $d=1\mu m$ \textbf{(a)}, and $d=2.8\mu m$ (b). Particle motion
occurs from left to right. In (a) the magnetic bubble in the shaded column requires a filling of $\tilde\rho>8$ particles to emit
particles towards the next bubble. In (b) the particle transport occurs
when the bubbles are filled with $\tilde\rho>1$ particles.
(c) Normalized current $\tilde{j}$ for particles of size $d=1 \mu m$, blue symbols, and of
$d=2.8 \mu m$, black symbols, as a function of the overloading, $\tilde\rho-N_c$
(d) Partially filled integer plateaus in the normalized current for $d=2.8 \mu m$ particles as a function of
the amplitude of $H_z$. (Movie {\it ad fig 2.AVI}).}
\label{figure2}
\end{center}
\end{figure}
The external field modulates the potential
and increases or decreases the potential wells of the magnetic bubbles
when it is parallel or antiparallel to the bubble magnetization.
Before the field is oriented completely antiparallel to the bubble, two additional energy wells per
unit cell nucleate in the marked blue and red interstitial regions in Fig.1(b). Due to
the parallel component of the field, $H_x$, the nucleation sites are displaced from the centre of the
three surrounding bubbles and are located in the proximity of the particle "emitter" bubble.
The preference for one of the three bubbles vanishes for completely antiparallel field orientations $H_x=0$, and
reverses when the parallel field orients mirror
\begin{figure*}
\begin{center}
\includegraphics[width=\textwidth,keepaspectratio]{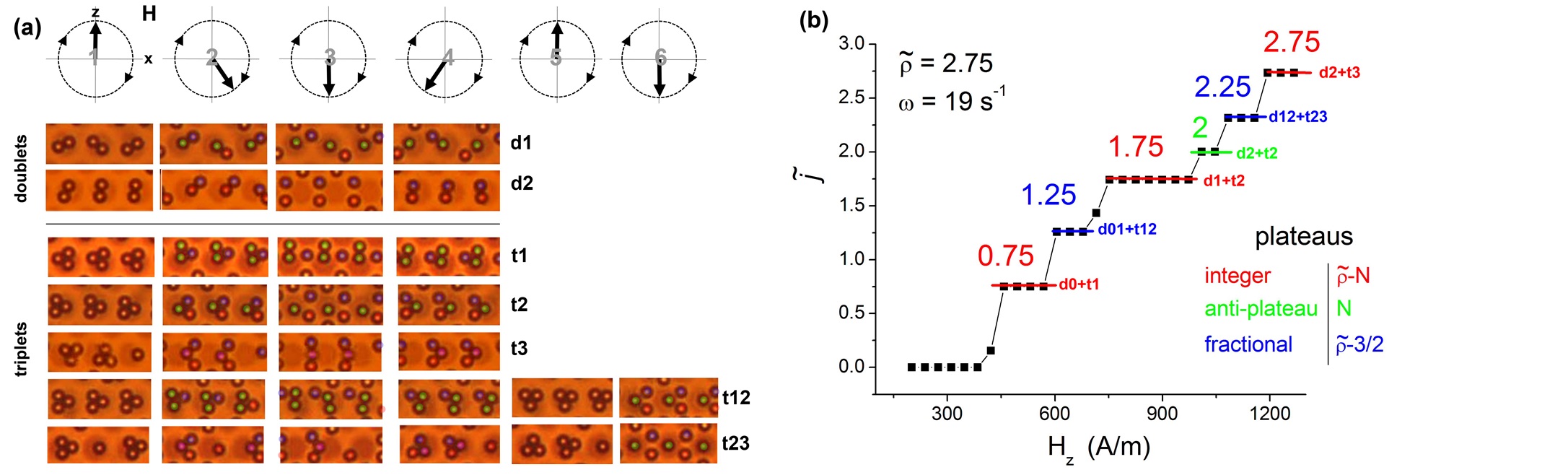}
\caption{(color online)(a) Polarization microscopy images showing
the time evolution of the doublet $d$ modes and triplet $t$ modes. The orientation of the external
magnetic field during the different stages is shown in the top row.
In the images where interstitial wells exist, the stored particles are marked
in green, particles in the type up or type down interstitial are marked in blue or red resp., and
particles hopping from annihilated emitter bubbles are marked in purple. The $t12$
and $t23$ modes are period doubled modes, where the second cycle (columns 5 and 6) differs from the first cycle.
(b) Dimensionless current $\tilde{j}$ as a function of $H_z$ for a
density of $\tilde\rho=2.75$ exhibiting different kind of plateaus:  partially filled
integer plateaus in red, fully filled integer plateaus in green, and partially filled
fractional plateaus in blue, together with the dominating modes (Movie {\it ad fig 3.AVI}).}
\label{figure3}
\end{center}
\end{figure*}
symmetrically $H_{x,annihil}=-H_{x,nucl}$ with respect to the orientation during the nucleation.
The interstitial wells hence nucleate near a bubble which  becomes an emitter of particles and they annihilate
near the collector bubble i.e. the neighboring bubble in transport direction.
The chirality of the external magnetic field modulation ensures unidirectional
motion from the emitter bubble toward the collector bubble.
A directed ratchet current is induced when a particle moves from an emitter
bubble into a collector bubble via the interstitial region.\\
The collective effect of the excluded volume on the transport for small ($1 \mu m$) particles is shown in Fig.2(a) (movie
{\it ad fig 2.AVI}).
Individual particles will reside in the bubble wells, while bubbles filled
with several particles generate a composite cluster
with a radius, $r_{cluster}$.
The green shaded column in Fig.2(a) separates the unloaded lattice, on the right,
from a series of loaded bubbles thereby forming composite clusters, on the left.
The particles are transported from the left bubbles
passing via the interstitial regions
to the marked
unfilled bubble, which they reach at $t=0.1s$,
and they start to grow a new composite cluster.
For small values of $r_{cluster}$,
this composite cluster is unable to emit particles into the interstitial wells to its right,
and no net current is observed.
For $t>1.35 s$, the cluster size
surmounts a critical value, $r_{cluster}>r_c \sim 2 \mu m$, and the particles located within
$r_{c}$ remain immobile, while excess particles are emitted to the right into an interstitial and transported to the next collector bubble.
In each cycle, the excess particles flow
via the interstitials, filling more bubbles
till reaching a stationary state that corresponds to
a particle loading of
$r_{cluster}\approx r_c$ for all bubbles.\\
We measure the current $\mathbf{j(\rho)}=\rho\mathbf{v}$
of particles with velocity
$\mathbf{v}$ as a function of the particle number area density $\rho$.
Figure 2(c) shows the dimensionless
current, $\tilde j=2\pi Aj/\omega a_1$ versus the dimensionless density $\tilde \rho= \rho A$,
where $A=a_1^2 sin(\pi/3)$ is the area of the unit cell, $a_1$ is the length of the unit vector in the $01$
direction, and $\omega$ is the angular frequency of the external field. We  observe no current below an average
loading of $\tilde\rho\le 8$ particles per unit cell. Only if $\tilde \rho >N_c=8$, a net current is observed, which
linearly increases with the density
as $\tilde j =\tilde \rho -N_c$. Eight particles in each unit cell remain immobile while only the excess particles
contribute to the current.\\
While small particles require the formation of highly populated clusters to
produce net flow, we can simplify the complexity of the transport
by increasing
the particle size. In Fig.2(b) we show the filling of a magnetic lattice using
large ($2.8 \mu m$) particles, which reduces the size of
the critical cluster to $N_c=0-4$ particles per unit cell. As a consequence, the filling process is much
faster than before, and the colloidal front propagates by one composite cluster every second cycle
rather than every fourth cycle.
The net current produced by the large particles follows the same law $\tilde j =\tilde \rho -N_c$
as the small particles, but with $N_c=1$, i.e. exactly
one particle per unit cell remains immobile while excess particles are transported.\\
In Fig. 2(d) we explore
the dependence of $\tilde{j}$ on the amplitude of the normal component of the applied field $H_z$,
for a fixed density $\tilde{\rho}=1.36$.
No current is observed for amplitudes $H_z < H^c_1= 780$ $A/m$. Beyond the threshold $H^c_1$,
only one particle per doublet is mobilized, while individual particles do not move, and the flux reaches
the constant plateau $\tilde{j}=0.36$.
Above a second threshold field, $H^c_2=1200$ $A/m$, all particles are mobilized,
and they
flow at a constant speed above the lattice.
The fields $H^c_1$ and $H^c_2$ are therefore the mobility edges where different
sub groups of particles start to move.
The plateaus in Fig.2(d) correspond to $N_c=2,1,0$
fully occupied immobile sites. The rest of the particles are mobile and consist of fully and
partially occupied sites contributing $\tilde j =\tilde \rho -N_c$ to the macroscopic current.
There are six small interstitial regions around each magnetic bubble which can be separated in two types, Fig.1(b,c).
Type up and type down interstitials are marked in blue and in red resp., and
are characterized by a neighboring bubble in the $-21$ and $2-1$ direction.
In Fig.2(a,b), there are equal numbers of particles
flowing via the type up and type down interstitials,
and the macroscopic current is unpolarized.\\
A rich variety of transport modes can be obtained by tuning the particle density and the
magnetic field parameters.
In Fig.3(a) (movie
{\it ad fig 3.AVI}) we show modes which arise for magnetic bubbles filled with doublets
($d$ modes) and triplets ($t$ modes). For field amplitudes $H_z<400 \, A/m$, no particles are transported corresponding to
the trivial $d0$ and $t0$ modes. The first modes giving rise to net particle current are the $d1$ and $t1$ mode.
The $d1$ mode occurs for fields in the range  $700  \, A/m <H_z<1100 \, A/m$ and is characterized by the
break up of a doublet into a stored particle in the bubble and an emitted particle in the interstitial.
The emitted particle is transported via type up, marked in blue,
or type down, marked in red, interstitials, and then collected by a magnetic bubble occupied by a single particle.
During the next cycle,
particles which were immobile during the first field cycle
become mobile, and vice versa.
The $t1$ mode starts with a particle triplet oriented with two corners toward
the interstitials. Before the interstitial wells are nucleated, the triplets undergo a rotation by $\pi/6$
followed by the emission of a single particle into one of the interstitial regions and the collection by a doublet with the final
formation of a triplet having the initial orientation.
In contrast, the $t2$ mode starts with a particle triplet oriented with one corner between the interstitials followed by a rotation by $\pi/6$
and an emission of two particles which simultaneously
follow both types of interstitials towards the collector
bubble.
The $d2$ and $t3$ modes are fully mobile transport modes
where all particles hop from the emitter to the collector bubble
and are found for magnetic fields higher than $H_z>1200 \, A/m$, that during the antiparallel orientation also annihilate the bubble wells.
The triplet rotation
takes more time than the actual particle hopping.
Thus, increasing the driving frequency allows us to freeze the rotational motion,
giving rise to two period doubling modes, $t12$ and $t23$.
Since the hopping inverts the orientation of the triplets,
two cycles are required to restore the original triplet orientation.
For both modes, the number of particles hopping during the first and the second cycles
are different, and this gives rise to an average fractional current of $3/2$ for
the $t12$ mode and $5/2$ for the $t23$ mode.\\
In Fig.3(b) we show measurements of $\tilde{j}$ as a function of the
field amplitude $H_z$ for a particle density ($\tilde{\rho}=2.75$) for which
the dominant colloidal clusters in the bubbles are doublets and triplets.
An increase of $H_z$ leads to a superposition of doublet and triplet modes
resulting in partially filled
integer plateaus ($\tilde{\rho}-N$), fully occupied integer plateaus  ($N$),
and fractional partially filled plateaus ($\tilde{\rho}-3/2$).
Fractional plateaus occur due to period-doubled modes where first an odd and secondly an even number of particles
are transported during the first and second field cycle.
Further experiments produced tetramers and highly ordered clusters
with even more complex transport modes and dynamics.\\
Our magnetic ratchet allows to create polarized colloidal
currents, using an additional field oriented along the $-21$ ($y-$direction).
We demonstrate this for the $d1$ mode in Fig.4(a,b), and for the $t1$ mode in Fig.4(c,d).
When using the rotating field in the $(x,z)$ plane,
both modes are characterized by a unpolarized current, with the emission
of one particle which randomly hops towards the collector bubble via
either the type up or type down interstitials.
The application of an alternating field along the $y-$direction,
$H_y=H_0 \sin{(\omega_y t)}$, phase-locked to the rotating field
with $\omega_y=\omega/2=9.4 \, s^{-1}$,
polarizes the particle emission in the $d1$ mode,
(Fig.4(b), movie {\it ad fig 4.AVI}).
With the $y$-field, the $d1$ mode carries a macroscopic alternating polarized current,
since the particles are periodically displaced in the type up and type down
interstitials, Fig.4(a).
\begin{figure}[t]
\begin{center}
\includegraphics[width=\columnwidth,keepaspectratio]{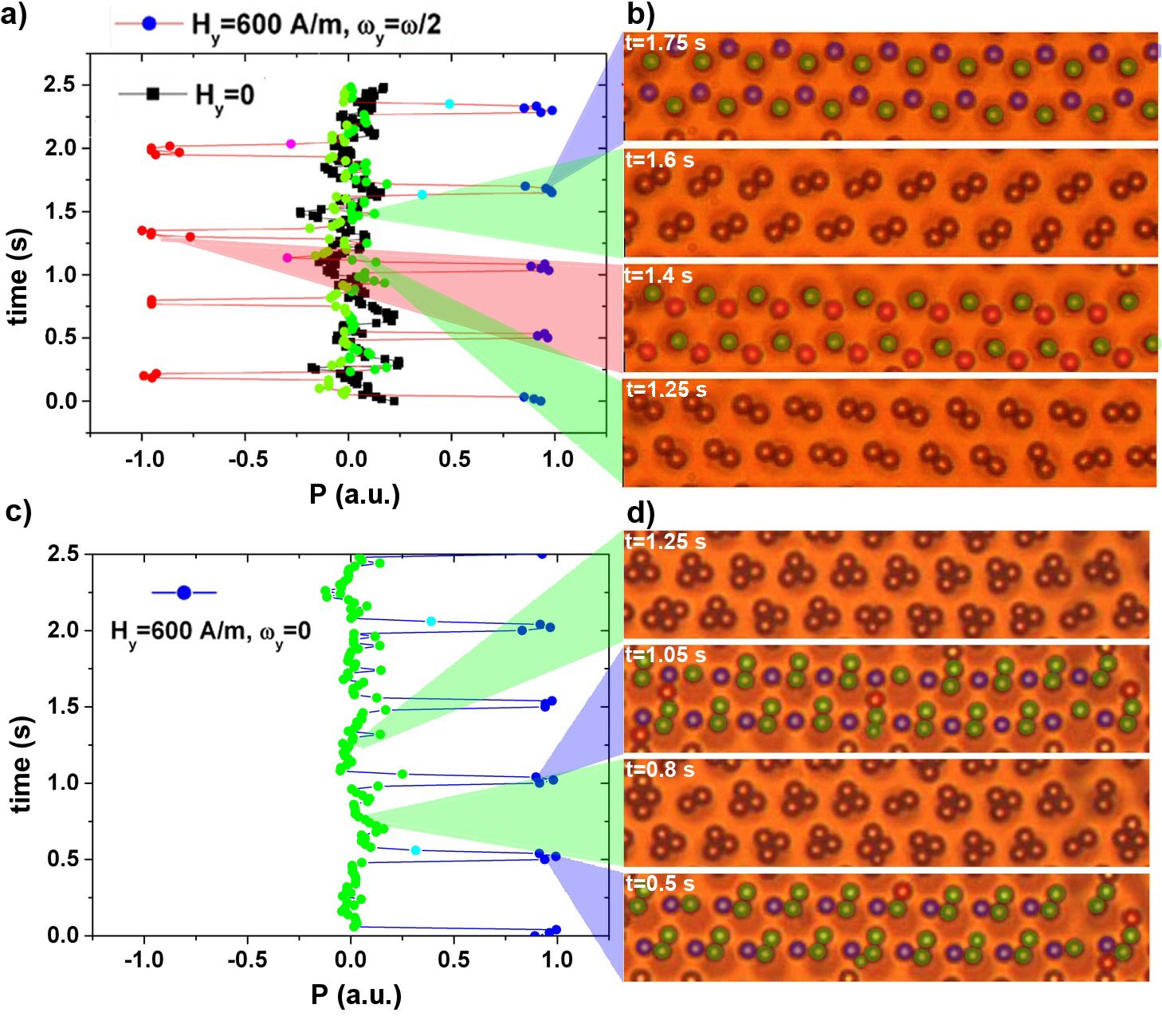}
\caption{(color online)(a) The polarization $P$
measured as a function of time without (squares)
and with (circles) an additional field along the $y-$direction for the $d1$ mode. (b) Four microscopy images corresponding to the first cycle doublet storage
($t=1.25 s$), hopping in type down interstitial ($t=1.4 s$), second cycle doublet storage ($t=1.6 s$),
and hopping in type up interstitial period ($t=1.75 s$). (c) The polarization per unit cell 
measured as a function of time with the field along the $y-$direction for the $t1$ mode.
(d) Four microscopy images corresponding to the first cycle hopping $t=0.5 s$,
triplet storage $t=0.8 s$, and a similar second cycle.
Hopping particles are marked in blue or red when moving into an interstitial of type up or down, while particles stored
in bubbles are marked in green, Movie {\it ad fig 4.AVI}.}
\label{figure4}
\end{center}
\end{figure}
We created fully polarized direct currents by increasing the particle density and accessing odd
triplet modes. We used a static field, $H_y=600 \, A/m$, $\omega_y=0$
to displace the triplet such that the corner of the triplet
before emission, lay close to the type up interstitial.
The resulting macroscopic current of the t1 mode displays an alternating polarization during each
half cycle, Fig. 4(c).
The alternating and direct polarization currents are collective effects only achieved with even and odd modes, resp.
The same principle is  applied to the $t12$ mode (not shown here):
A $y$-field induces polarized hopping of single particles via type up interstitials during the first field cycle,
while unpolarized hopping occurs during the second cycle when
two particles are emitted to both the type up and down interstitials.
This realizes a macroscopic fractional current
since the net current flowing in the type up interstitial transports one particle per cycle while
the current in the type down transports only half a particle per cycle.\\
In summary, our experiments show that excluded volume between mesoscopic
particles gives rise to ratchet transport modes where $n$ particle steps occur during a
period consisting of $m$ cycles of the field, contributing with $\tilde{j} = n/m$ to the particle current.
For integer particle filling of the bubbles, only a single mode is selected. 
If the filling is incommensurate with the bubble lattice, a superposition of
modes is observed due to the
inhomogeneous distribution of composite clusters across the bubbles.
The total current is the sum of the currents associated with each transport mode and remains at simple integer or fractional plateaus.
The mobility or immobility of the partial particle layer determines whether one has a partially filled or fully filled plateau in the current.\\
We thank Tom H. Johansen for the FGF film and
Matthias Schmidt for scientific discussions.
P.T. acknowledges support from the ERC starting grant "DynaMO" (335040)
and from the programs RYC-2011-07605, and FIS2011-15948-E.

\end{document}